\documentclass{ws-p9-75x6-50}
\usepackage{epsfig}
\begin{document}
\title{A study of the gravitational wave form from pulsars}
\author{S. R. Valluri}
\address{Department of Physics \& Astronomy, University of Western Ontario, London,
Ontario
N6A 5B7, Canada\\Email: valluri@julian.uwo.ca }
\author{F. A. Chishtie}
\address{Department of Applied Mathematics, University of Western Ontario, London,
Ontario
N6A 5B7, Canada\\Email: fachisht@julian.uwo.ca }
\author{R. G. Biggs}
\address{Department of Mathematics, University of Western Ontario, London, Ontario N6A
5B7, Canada\\Email: rbiggs@julian.uwo.ca }
\author{M. Davison}
\address{Department of Applied Mathematics, University of Western Ontario, London,
Ontario
N6A 5B7, Canada\\Email: mdavison@julian.uwo.ca }
\author{Sanjeev V. Dhurandhar}
\address{Inter-University Centre for Astronomy and Astrophysics, Post Bag 4, Ganeshkhind ,
Pune 411 007, India\\Email: sdh@iucaa.ernet.in }
\author{B. S. Sathyaprakash}
\address{Department of Physics and Astronomy, University of Cardiff, Cardiff, UK\\Email:
b.s.sathyaprakash@astro.cf.ac.uk }
      
\maketitle \abstracts {We present analytical and numerical studies of the 
Fourier transform (FT) of the gravitational wave (GW) signal from a pulsar,
taking into account the rotation of the Earth for a  one day observation 
period. }

\section{INTRODUCTION }
The direct detection of gravitational radiation (GR) from astrophysical
sources is one of  the most important
outstanding problems  in experimental
gravitation
today. The construction of large laser 
interferometric gravitational wave detectors like the LIGO\cite{Abr92},
VIRGO \cite{jot96}, 
LISA, TAMA 300,
GITO 6000 and AIGO
is opening
a new  window for the study of a vast and  rich variety of
nonlinear curvature phenomena.
The network of gravitational wave detectors 
can confirm that GW exist and by monitoring gravitational
wave forms give important information  on their amplitudes, frequencies 
and other important physical parameters. 

A prototype of continuous astrophysical sources is a pulsar. A variety of
instabilities cause
deformations from spherical
symmetry giving rise to GWs. The amplitudes of GR from these pulsars
are probably very
weak($ \le 10^{-26} - 10^{-28}$, for galactic pulsars).
The GR signal will be buried deep
within
the noise of the detector system.  The detection of a GR signal 
warrants the urgent need of careful data analysis with development of
{\em analytical methods} and {\em  problem oriented algorithms.}

In sections 2, 3 and 4 we outline the approach leading to the
FT of the GW signal.
The frequency modulation (FM),  Doppler shift due to rotation and orbital
motion of the Earth in the 
Solar System Barycentre (SSB) frame, its effect on the total phase of the 
received GW signal  
and the Fourier transform (FT) of the GW signal have
previously been described\cite{jot96}.

 Sections 5 and 6 present discussion and conclusions.

\section{Methodology}
\label{pulsign}
\smallskip

Typical values of the gravitational wave amplitude
$h$ for the Crab and Vela pulsars are 
$\sim 10^{-25}$ and $ \sim 10^{-24}$ respectively. 
This amplitude is several  orders 
of magnitude below LIGO's expected sensitivity of $\sim 10^{-23}$).
Since the LIGO would make continuous
observations over a time scale of a few
months  or more, a significant enhancement
to the signal-to-noise
ratio (SNR) is expected by integrating the data over a long time interval.

The total response of the detector is a
function of the source position,
the detector orientation,
the orientation of the spin axis of
the Earth and the orientation of the orbital plane.  Since the pulsar  signal
is weak, 
long integration times $\approx 10^{7}$ secs  will be needed to extract the 
signal from the noise. Since the detector moves along
with the Earth in
this time, the frequency of the wave emitted by the source is Doppler 
shifted. Also since the detector has an anisotropic response, the signal 
recorded by the detector is both frequency and amplitude modulated.
We  discuss now the important role of
frequency modulation in the context of signal
detection.

\section{Study of   Frequency Modulated Pulsar Signal}
\label{discrifm}
Frequency modulation  arises due to translatory motion of the detector
acquired from the motion of the Earth. 
We consider only two motions of
the Earth:  its  rotation about the   
 spin axis and the  orbital motion about the Sun, so the response is doubly 
frequency modulated with one period corresponding to a day and the other 
period corresponding to a year.
 The FM smears out  a 
monochromatic signal into a small  bandwidth around
the signal  frequency of the
monochromatic 
waves. It also redistributes the power in a  small bandwidth.  
The study of FM due to rotation  of the Earth 
about its spin axis for a one day observation period
shows that  the Doppler spread
in the  angular 
bandwidth for 1kHz signal will be 0.029Hz. The Doppler spread in the 
angular bandwidth due to  
orbital motion for an observation period of one day
will be $1.7 \times 10^{-3}$Hz
\cite{jot96}. 

Since any observation is likely to last longer than a day, {\em it 
is important to incorporate this effect in the data analysis 
algorithms. }

In order to study frequency modulation of a monochromatic plane  wave, 
one must calculate the Doppler shift due to rotational
and orbital motion of the Earth in
the SSB frame. For this, we need to know  relative velocity between the  
source and  detector.
The Euler angles  $ \left(\theta, \phi\right) $ give
the direction of the  incoming
wave in the   SSB frame. We characterize the motion of the Earth (and detector)
simply:  (a) We assume the orbit of the Earth to be circular. 
  (b) We neglect the effect of the Moon on the motion of the Earth.

The phase $\phi(t)$ of the Doppler shifted received signal for a single direction sky search 
$(\theta, \phi)$ is given by, 
\begin{eqnarray}  
 \phi (t) & = & 2 \pi   \int_{t_{\scriptscriptstyle 0}}^{t} f_{rec} 
 \left(t^{\prime}\right) d t^{\prime} \\
 & = & 2 \pi  {f}_{\scriptscriptstyle 0} \int_{t_{\scriptscriptstyle 0}}^{t} 
 \left( 1 + { {\vec v} \cdot {\vec n} \over c }\left(t^{\prime}\right) \right)
 d t^{\prime} \\
& = & 2 \pi {f}_{ \scriptscriptstyle 0}  \Bigg[ t -t_{ \scriptscriptstyle 0} +
\bigg\{ { A \over c} \sin\theta 
\cos\phi^{\prime} + { R \over c} \sin\alpha \big\{ \sin\theta    
 \big( \sin\beta^{\prime} \cos\varepsilon \sin\phi \nonumber \\ 
& &   + \cos \phi \cos \beta^{\prime} \big)  
 + \sin\beta^{\prime} \sin \varepsilon \cos \theta \big\} \bigg\}  
 - \bigg\{  {A \over c} \sin\theta \cos \phi^{\prime}_{ \scriptscriptstyle 0}  
 + { R \over c} \sin\alpha \nonumber \\
& &   \big\{ \sin\theta \big(
\sin\beta^{\prime}_{\scriptscriptstyle 0} \cos\varepsilon \sin\phi  
 + \cos \phi \cos \beta^{\prime}_{ \scriptscriptstyle 0} \big)
 + \sin\beta^{\prime}_{\scriptscriptstyle 0} \sin \varepsilon \cos \theta  \big\} \bigg\}
\Bigg] . \nonumber \\  \label{doppphdel}    
\end{eqnarray}
where  $ \phi^{\prime} = \omega_{orb} t - \phi,  \ 
\beta^{\prime} = \beta_{\scriptscriptstyle 0} + \omega_{rot} t, \ \phi^{\prime}_{
\scriptscriptstyle 0} = \omega_{orb} t_{\scriptscriptstyle 0 } - 
\phi, \  \beta^{\prime}_{\scriptscriptstyle 0}  = \beta_{\scriptscriptstyle 0} 
+ \omega_{rot} t_{\scriptscriptstyle 0},  \  \beta_{\scriptscriptstyle 0} $  
is the initial azimuthal angle of the detector at the observation time 
$t_{\scriptstyle 0}. $ 
where $A$ is distance from the 
centre of the SSB frame to the centre of the Earth, $R$ is the radius of
the Earth,
and ${\vec n}$ the  unit vector in the direction of source,   
$ {\vec n} \, = \, \big( \sin\theta \cos\phi, \,
\sin\theta\sin\phi, \, \cos\theta \big).$  Here we have assumed that at time 
$t =0,$ the longitudinal angle $\beta =0.$

It can be seen  from eq~(\ref{doppphdel}) that the Doppler 
corrections to the phase of received pulsar signal depends on the direction 
of the source in  the sky.
Orbital motion is omitted in this preliminary analysis but will
be presented in later work.

\section{Fourier Transform Analysis  of the FM signal due to the 
Rotational Motion of the Earth}   
We analyze the Fourier transform (FT) of the frequency 
modulated signal and study the extent to which the peak of the FT is  smudged 
and to how much  the FT {\em spreads} in the frequency space. This type of 
study would be useful from the point of view of data analysis and 
for applying  such schemes as {\em stepping  around the sky} method 
\cite{Schutz91} which relies on the FT.

We let ${\scriptstyle X} = {2 \pi {f}_{\scriptscriptstyle 0}{\scriptstyle R}
\over c}$ and $t_{\scriptscriptstyle 0} = 0.$
Here  ${\scriptstyle X}$ plays the role
of modulation index similar to  $K$ in
the theory of signal modulation. The
modulation index depends on the frequency of the
pulsar signal.
If we consider only the  frequency modulated output of the signal, the 
output of amplitude unity is given as follows, 
\begin{equation} 
h \left(t\right) \, = \, \cos \left( \phi\left(t\right)\right). 
\label{htsig}
\end{equation}  
We now consider the $h\left(t\right)$ to be given on a finite time interval 
$[ 0, T]$ which would be assumed to be the
observation period. In our analysis we
have assumed $T$ to be one day. 
The Fourier transform of the signal $h \left(t\right)$ is given by,
\begin{equation}
{\tilde h} \left(f\right) \, = \, \int_{0}^{T} h \left(t\right) 
e^{ - \,  i 2  \pi f  t} d t.  \label{ftsig}
\end{equation}  
It is convenient to use a time coordinate  $\xi  = {\omega}_{rot} t/2$ 
which  for a period of a day is of the order of unity, {\em i.e} when 
$T = 1$day $ = 86400$ secs,  the ${\xi}_{\scriptscriptstyle T} 
= {\omega}_{rot} T = \pi .$

An exact closed form  expression for the Fourier
transform of the  frequency modulated GW
signal is obtained by the analytical approach.
The plane wave expansion in spherical harmonics
is used. \cite{Brans90}

Then the GW signal becomes 
$$\sum_{l,m}\int  S_{lm} (f_0,\alpha,t)dt =$$
\begin{equation}
 \Re\int 4\pi [\sum_{\rho =0}^\infty
i^lY_{lm}^\ast(\theta,\phi)e^{2\pi i f_0t}
j_l(k\sin(\omega_rt/2))Y_{lm}(\alpha,{\omega_rt\over 2})] dt
\end{equation}

where $\bar K$ is the wave vector with
spherical coordinates $(K,\theta,\phi)$ with
\begin{equation}
|K|=4\pi f_0\Re \sin \alpha (\sin (\omega_rt/2)) = k\sin(\omega_rt/2)
\end{equation}

and $\hat n= (n,\theta,\phi)$ ($\hat n=1$).

and the FT is given by
\begin{equation}S_{lm}(f,f_0,\alpha)=FT(S_{lm}(f_0,\alpha,t))\end{equation}
Further simpification gives:
\begin{equation}\sum _{lm} S_{lm}(\omega,\omega_0,\alpha)=\sum_l\sum_m
A_{lm}I(l,B,k)\end{equation} 

where 

\begin{equation}A_{lm}={8\over \sqrt{2}}\pi^{3/2}i^lN_{lm}P_l^m(\cos
\alpha)Y_{lm}^\ast
(\theta,\phi)\end{equation} 

and 
\begin{equation}N_{lm}=\sqrt {{(2l+1)(l-|m|)!}\over {4\pi (l+|m|)!}}\end{equation}

With convenient co-ordinate transformations, we obtain  
\begin{equation}S_{lm}(\omega ,\omega_0,\alpha)=A_{lm}I(l,B,k)\end{equation}
 where
\begin{equation}I(l,B,k)=\int_0^\pi \exp [iB\xi ] {J_{l+1/2}(k \sin\xi )\over \sqrt{k\sin \xi}}
d\xi\end{equation}
Also, 
\begin{equation}I(l,B,k)=2i^B\int_0^1 {{\cos(B\cos^{-1} x)J_{l+1/2}(kx)}\over
{\sqrt{kx}\sqrt{1-x^2}}}   dx\end{equation}

(where $B=2({\omega_0 -\omega\over \omega_r}) +m$ )

which equals

\begin{equation}2\pi{i^B\over\sqrt{2}}{k^{l+1/2}\over2^{2l+1}}{\Gamma(l+1)\over
\Gamma(l+3/2)}
{{_1F_3(l+1;l+3/2,(l+2+B)/2,(l+2-B)/2;-k^2/16)}\over
{\Gamma((l+2+B)/2)\Gamma((l+2-B)/2)}}\end{equation}

The $_1F_3$ is the generalised
hypergeometric function with one parameter (here $l+1$) in
the 
numerator and
three parameters in the denominator and $-k^2/16$ as its variable.
A wealth of information is hidden
in these parameters and $_1F_3$ is uniformly convergent.

The numerical integration is done
via the Clenshaw-Curtis quadrature algorithm and gives
accurate and reliable results. Both the analytical
and numerical results agree for $k<1$ and
arbitrary $l$; a result of possible relevance
for LISA which observes the low frequency band
0.1 to 0.00001 Hz. For simultaneously large
$k$ and $l$, the factor $k^{(l+1/2)}$ presents a
challenge.

We also evaluate  
\begin{equation}|S_{lm}(K)|^2 \propto {\pi\over 4} {k^{2l}\over 2^{2l+1}}
{{\Gamma(1/2)\Gamma(l+1/2) _1F_2(l+1/2;2l+2,l+3/2;-k^2)}\over {\Gamma(l+1)
\Gamma(l+3/2)\Gamma(l+3/2) }}\end{equation}

In this case we find perfect agreement
with numerical results for $k\leq 86$ and arbitrary $l
(\leq 300)$.
The above approach using the
plane wave representation appears to be equally useful when
orbital
motion is also included in the
analysis and will be discussed in future work involving orbital
corrections.

\section{Discussion}
In general, the FT is a complex valued function of the 
parameters $k$, $l$ and $B$.
Where appropriate, the absolute, real and the imaginary parts
are
taken.  In Figure 1, the absolute value of the
integral is plotted as a function of $k$
for $B=30$ and $l=10$. The maximum occurs around
$k=40$, 
which translates to approximately 150 Hz. The intensity drops off fast and is negligible
around
$k=300$. Figure 2 again shows the absolute value of $I$ vs $k$ for $B=10$
but for $l=270 >> 1$.
This plot shows remarkable evidence for the Bessel function index theorem
as it shows $I(k) = 0$ for $k<250$, rising to a
maximum around $k=285$ and falling off fast thereafter.
For both large $k$ and $l$, $B$ must be
small to control 
the oscillations of the integrand to give
a non-negligible value. Figure 3 shows absolute $I$ vs
$B$ for $l=k=10$.
The integral peaks at $B=0$ and drops off
fast around $B=6.25$ and rises again to a tiny
value before dropping
off to zero $B\geq10$.
Figure 4
shows absolute $I$ vs $B$ for $l=k=200 (>>1)$. The
integral
peaks at $B=0$ then drops off rapidly at $B>20$.
This again illustrates that $B$ has to be
small
to control the oscillations.
Figure 5 shows absolute $I$ vs $B$ for
$l=10$ and $k=270 (>>1)$. Here the many bands of
$B$ are in full display until $B\leq 280$.
Figure 6 shows the absolute value
$|S_{lm}(k)|^2$ vs
$l$ and $k$ for the indicated range.

  \begin{figure}[htb]
       \epsfig{file=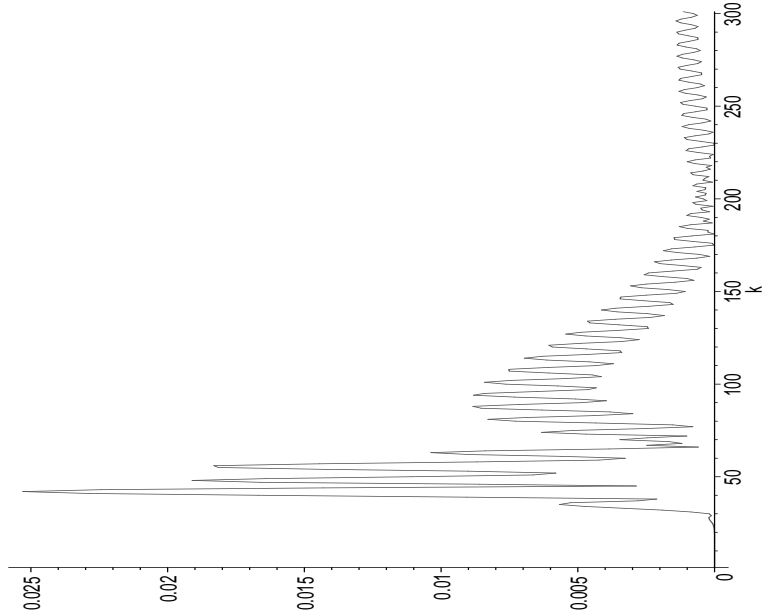, width=10cm, height = 8cm} 
       \caption{$|I(l,k,B)|$ vs $k$ for $B=30, l=10$}
  \end{figure}

\begin{figure}[htb]   

    \epsfig{file=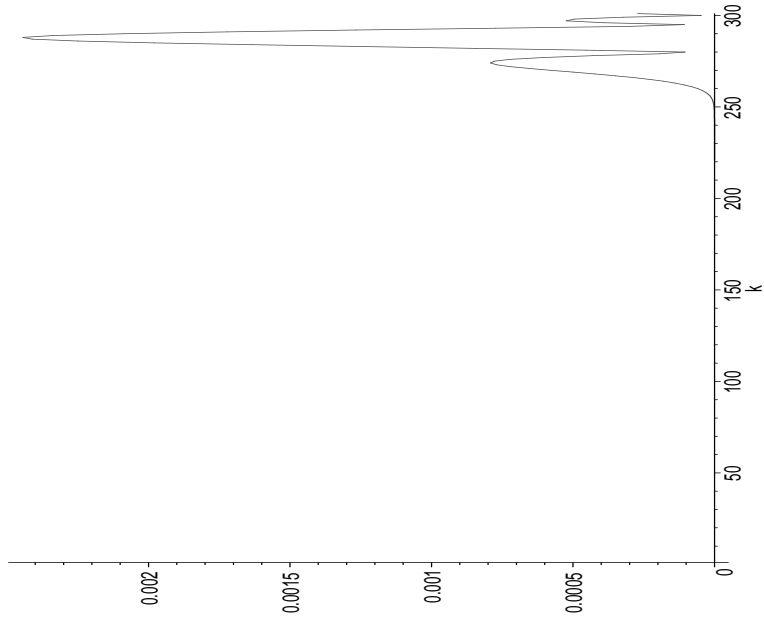, width = 10cm, height=8cm}
    \caption{$|I(l,k,B)|$ vs $k$ for $B=10, l=270$, width=10cm, height=9cm}
 
\end{figure}     
 
\begin{figure}[htb]   
 
    \epsfig{file=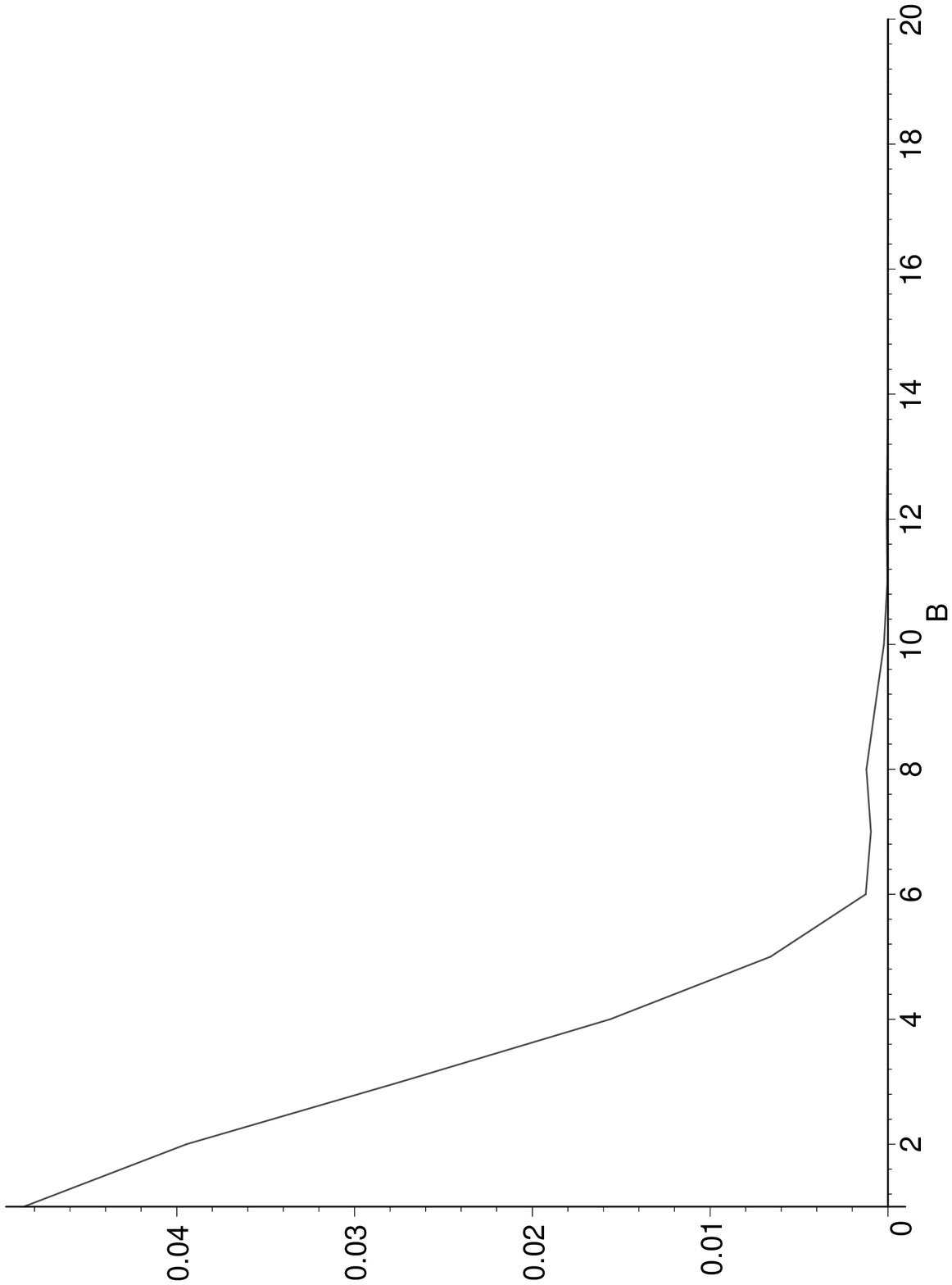, width = 10cm, height = 8cm}
    \caption{$|I(l,k,B)|$ vs $B$ for $k=10, l=10$, width=10cm, height=9cm}
 
\end{figure} 

\begin{figure}[htb]   
 
    \epsfig{file=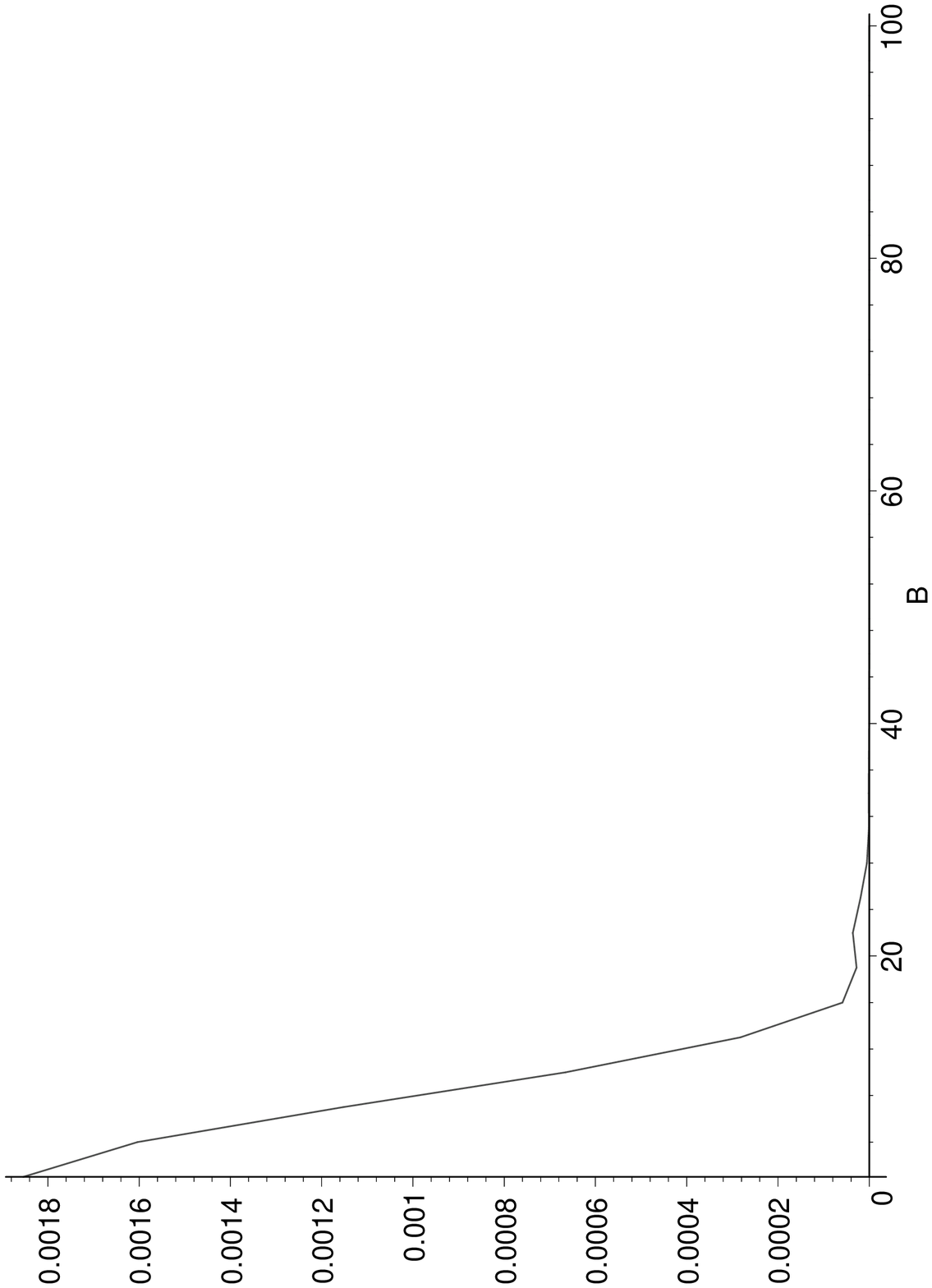, width = 10cm, height = 8cm }
    \caption{$|I(l,k,B)|$ vs $B$ for $k=200, l=200$}
 
\end{figure} 
 
\begin{figure}[htb]   
  
    \epsfig{file=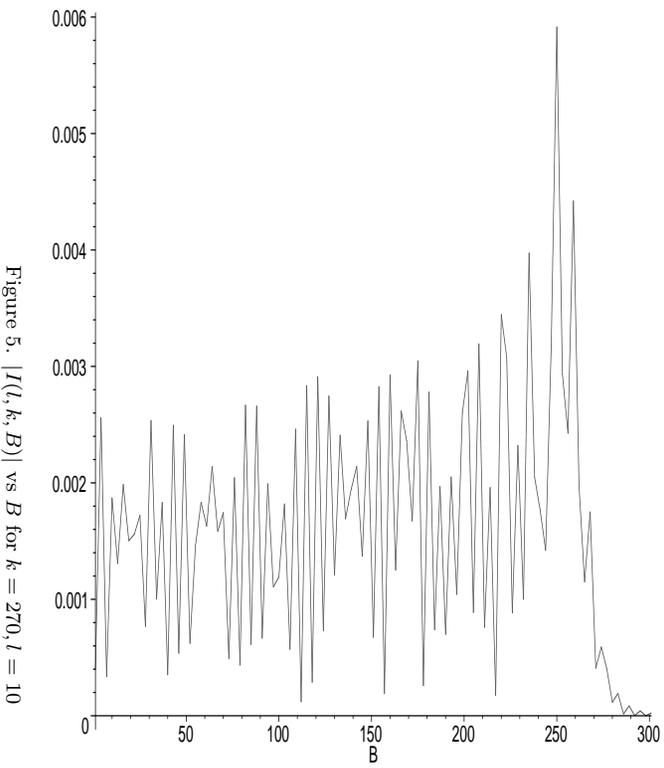, width=10cm, height=8cm}
    \caption{$|I(l,k,B)|$ vs $B$ for $k=270, l=10$}
 
\end{figure} 
 
\begin{figure}[htb]   
  
    \epsfig{file=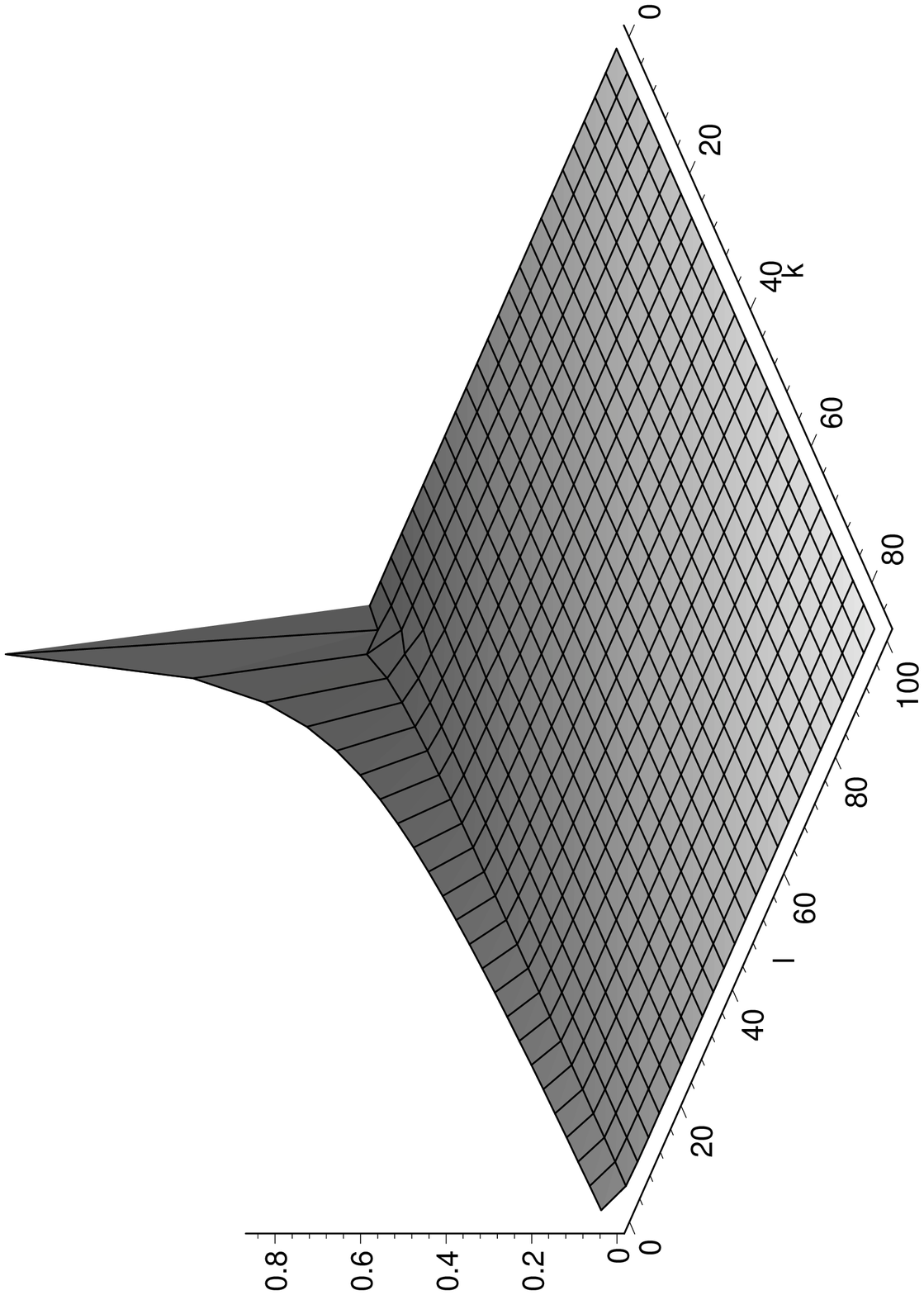, width=10cm, height = 8cm}
    \caption{$|S_lm(k)|^2$ vs $k$ and $l$ for $0<k<86$, $0<l<100$}
 
\end{figure}

\section{Conclusion} 
\label{concl}
We have studied the frequency 
modulation of the pulsar signal by two different methods.  A
closed  form of the Fourier transform of the frequency modulated GW pulsar 
signal due to rotational motion of the Earth about
its spin axis has been obtained. The work
with
the inclusion of orbital corrections results
in a double series with only a finite number of
surviving terms. Further details of this
analysis warrant study and will be presented
in later work\cite{Valet}.   Deeper results would
be useful for schemes like the stepping
around
the sky method\cite{Schutz91} and for differential
geometric methods that allow the setup
of
search templates in the relevant parameter space\cite{Sathy98}.
This analysis should
also be applicable to the Doppler
modulation of the GW signal caused by LISA's orbit around
the sun.

\end{document}